\documentclass[11pt,a4paper]{article}
\usepackage{jheppub}
\usepackage{amsmath,amssymb,graphicx,epstopdf,enumerate,amsfonts,booktabs,color}
\usepackage{float,placeins}
\usepackage{amsmath}
\usepackage{amsfonts}
\usepackage{amssymb}
\usepackage{graphicx}

\providecommand{\U}[1]{\protect\rule{.1in}{.1in}}
\affiliation[a]{Department of Electrophysics, National Chiao Tung University, Hsinchu, Taiwan, ROC}
\affiliation[b]{Department of Physics, Kangwon National University, Chuncheon 200-701 Korea}
\emailAdd{xgcj944137@gmail.com}
\emailAdd{jcclee@cc.nctu.edu.tw}
\emailAdd{taejin@kangwon.ac.kr}
\emailAdd{yiyang@mail.nctu.edu.tw}
\abstract{We show that there exist infinite number of recurrence relations valid for
\textit{all} energies among the open bosonic string scattering amplitudes
(SSA) of three tachyons and one arbitrary string state, or the Lauricella SSA.
Moreover, these infinite number of recurrence relations can be used to solve
all the Lauricella SSA and express them in terms of one single four tachyon
amplitude. These results extend the solvability of SSA at the high energy, fixed angle scattering limit and those at the Regge scattering limit discovered previously to all kinematic regimes.}
\input{tcilatex}
\begin{document}

\title{Solving Lauricella String Scattering Amplitudes through Recurrence
Relations}
\author{Sheng-Hong Lai${^{a}}$, Jen-Chi Lee${^{a}}$, Taejin Lee${^{b}}$, Yi
Yang${^{a}}$}
\maketitle

\setcounter{equation}{0}
\renewcommand{\theequation}{\arabic{section}.\arabic{equation}}%

\section{\protect\bigskip Introduction}

It was widely believed that string theory consists of huge spacetime
symmetries. In particular, it was conjectured \cite{GM,Gross,GrossManes}
that in the high energy fixed angle regime there existed infinite number of
\textit{linear} relations among string scattering ampitudes (SSA) of
different string states. Moreover, these relations are so powerful that they
can be used to solve all high energy SSA and express them in terms of one
single four tachyon amplitude.

Since it is a nontrivial task to compute exactly infinite number of massive
higher spin SSA and the corresponding exact symmetries \cite%
{Moore,Moore1,CKT} for states at general mass levels, only SSA at high
energy limit and relations among them were worked out in details in the
literature \cite{ChanLee1,ChanLee2,CHL,PRL, CHLTY,susy} \cite{LY,LY2014}. In
fact, in the high energy fixed angle regime, the existence of these infinite
linear relations among SSA was conjectured by Gross \cite%
{GM,Gross,GrossManes} and later corrected and explicitly proved in \cite%
{ChanLee1,ChanLee2,CHL,PRL, CHLTY,susy} by using decoupling of zero-norm
states (ZNS) \cite{ZNS1}, and can be used to reduce the number of
independent hard SSA from $\infty$ down to $1$.

On the other hand, it was discovered that in the high energy fixed momentum
transfer regime, the Regge SSA of three tachyons and one arbitrary string
states can be expressed in terms of the first Appell function $F_{1}$ \cite%
{LY2014}. Regge stringy symmetries or recurrence relations, instead of
linear relations in the fixed angle regime, among Regge SSA \cite{LY,LY2014}
were constructed and can be used to reduce the number of independent\ Regge
SSA from $\infty$ down to $1$. Moreover, an interesting link between Regge
SSA and hard SSA was identified \cite{KLY,LYY}, and for each mass level the
ratios among hard SSA can be extracted from the corresponding Regge SSA. See
the recent review \cite{review} for more details.

Recently a class of stringy SSA valid for \textit{all} energies were exactly
calculated \cite{LLY2}. These are the open bosonic SSA of three tachyons and
one arbitrary string state or the so-called Lauricella string scattering
amplitudes (LSSA). These LSSA were soon calculated by using the deformed
cubic string field theory \cite{TL,TL2,TL3,TL4}, and exactly the same
results were reproduced \cite{LLYL} consistently. This string field theory
calculation \cite{LLYL} can be considered as the extention of SSA
calculation of low mass string states \cite{Gidd,Samuel} through second
quantized string to those of infinite number of higher mass string states.
One important follow-up question then is whether there exist relations among
these LSSA so that one can use them to solve all the LSSA and express them
in terms of one single four tachyon amplitude.

In this paper, we will show that all the LSSA calculated in \cite{LLY2} can
be solved through various recurrence relations of Lauricella functions.
Moreover, all the LSSA can be expressed in terms of one single amplitude,
say four tachyon string scattering amplitude. These results strongly suggest
the existence of a huge spacetime symmetry of open bosonic string theory
associated with $SL(K+3,C)$ \cite{sl5c,sl4c} as suggested in \cite{LLY2},
and are extentions of results calculated previously for both high energy,
fixed angle SSA and Regge SSA \cite{review}.

In the last section of this paper, we will study the Lauricella zero norm
states (LZNS) and the corresponding stringy Ward identities. In particular,
we will show that the solvability of LSSA through recurrence relations imply
the validity of Ward identities. However, the LZNS or the corresponding
Lauricella Ward identities are \textit{not} good enough to solve all the
LSSA and express them in terms of one amplitude. This is in constrast to the
fact that the high energy zero norm states (HZNS) at the fixed angle regime
can be used to solve all the high energy SSA \cite%
{ChanLee1,ChanLee2,CHL,PRL, CHLTY,susy} and express them in terms of one
single four tachyon amplitude.

\section{The Lauricella String Scattering Amplitude}

We first briefly review the calculation of the LSSA \cite{LLY2} of three
tachyons and one arbitrary string states sitting at the second vertex. In
the center of momentum frame, the kinematics are defined as
\begin{subequations}
\begin{align}
k_{1} & =\left( \sqrt{M_{1}^{2}+|\vec{k_{1}}|^{2}},-|\vec{k_{1}}|,0\right) ,
\\
k_{2} & =\left( \sqrt{M_{2}+|\vec{k_{1}}|^{2}},+|\vec{k_{1}}|,0\right) , \\
k_{3} & =\left( -\sqrt{M_{3}^{2}+|\vec{k_{3}}|^{2}},-|\vec{k_{3}}|\cos
\phi,-|\vec{k_{3}}|\sin\phi\right) , \\
k_{4} & =\left( -\sqrt{M_{4}^{2}+|\vec{k_{3}}|^{2}},+|\vec{k_{3}}|\cos
\phi,+|\vec{k_{3}}|\sin\phi\right)
\end{align}
where $M_{1}^{2}=M_{3}^{2}=M_{4}^{2}=-2$ and $\phi$ is the scattering angle.
The Mandelstam variables are $s=-\left( k_{1}+k_{2}\right) ^{2},~t=-\left(
k_{2}+k_{3}\right) ^{2},~u=-\left( k_{1}+k_{3}\right) ^{2}$. The three
polarization vectors on the scattering plane are \cite{ChanLee1,ChanLee2}
\end{subequations}
\begin{subequations}
\begin{align}
e^{T} & =(0,0,1), \\
e^{L} & =\frac{1}{M_{2}}\left( |\vec{k_{1}}|,\sqrt{M_{2}+|\vec{k_{1}}|^{2}}%
,0\right) , \\
e^{P} & =\frac{1}{M_{2}}\left( \sqrt{M_{2}+|\vec{k_{1}}|^{2}},|\vec{k_{1}}%
|,0\right) .
\end{align}
Note that SSA of three tachyons and one arbitrary string state with
polarizations orthogonal to the scattering plane vanish. For later use, we
define
\end{subequations}
\begin{equation}
k_{i}^{X}\equiv e^{X}\cdot k_{i}\text{ \ for \ }X=\left( T,P,L\right) .
\end{equation}

For the $26D$ open bosonic string, the general string states at mass level $%
M_{2}^{2}=2(N-1)$, $N=\sum_{n,m,l>0}\left( np_{n}+mq_{m}+lr_{l}\right) $
with polarizations on the scattering plane are of the form%
\begin{equation}
\left\vert p_{n},q_{m},r_{l}\right\rangle =\prod_{n>0}\left( \alpha
_{-n}^{T}\right) ^{p_{n}}\prod_{m>0}\left( \alpha _{-m}^{P}\right)
^{q_{m}}\prod_{l>0}\left( \alpha _{-l}^{L}\right) ^{r_{l}}|0,k\rangle .
\label{p}
\end{equation}%
The $(s,t)$ channel of the LSSA were calculated to be \cite{LLY2}%
\begin{align}
A_{st}^{(p_{n};q_{m};r_{l})}& =B\left( -\frac{t}{2}-1,-\frac{s}{2}-1\right)
\prod_{n=1}\left[ -(n-1)!k_{3}^{T}\right] ^{p_{n}}  \notag \\
& \prod_{m=1}\left[ -(m-1)!k_{3}^{P}\right] ^{q_{m}}\prod_{l=1}\left[
-(l-1)!k_{3}^{L}\right] ^{r_{l}}  \notag \\
& \cdot F_{D}^{(K)}\left(
\begin{array}{c}
-\frac{t}{2}-1;\left\{ -p_{1}\right\} ^{1},\cdots ,\left\{ -p_{n}\right\}
^{n},\left\{ -q_{1}\right\} ^{1},\cdots ,\left\{ -q_{m}\right\} ^{m} \\
,\left\{ -r_{1}\right\} ^{1},\cdots ,\left\{ -r_{l}\right\} ^{l};\frac{u}{2}%
+2-N; \\
\left[ 1\right] ,\cdots ,\left[ 1\right] ,\left[ \tilde{z}_{1}^{P}\right]
,\cdots ,\left[ \tilde{z}_{m}^{P}\right] ,\left[ \tilde{z}_{1}^{L}\right]
...,\left[ \tilde{z}_{l}^{L}\right]
\end{array}%
\right)   \notag \\
& =B\left( -\frac{t}{2}-1,-\frac{s}{2}-1\right) \prod_{n=1}\left[
-(n-1)!k_{3}^{T}\right] ^{p_{n}}\prod_{m=1}\left[ -(m-1)!k_{3}^{P}\right]
^{q_{m}}\prod_{l=1}\left[ -(l-1)!k_{3}^{L}\right] ^{r_{l}}  \notag \\
& \cdot F_{D}^{(K)}\left( -\frac{t}{2}-1;R_{n}^{T},R_{m}^{P},R_{l}^{L};\frac{%
u}{2}+2-N;\tilde{Z}_{n}^{T},\tilde{Z}_{m}^{P},\tilde{Z}_{l}^{L}\right)
\label{st}
\end{align}%
where we have defined $R_{k}^{X}\equiv \left\{ -r_{1}^{X}\right\}
^{1},\cdots ,\left\{ -r_{k}^{X}\right\} ^{k}$ with $\left\{ a\right\} ^{n}=%
\underset{n}{\underbrace{a,a,\cdots ,a}}$, $Z_{k}^{X}\equiv \left[ z_{1}^{X}%
\right] ,\cdots ,\left[ z_{k}^{X}\right] $ with $\left[ z_{k}^{X}\right]
=z_{k0}^{X},\cdots ,z_{k\left( k-1\right) }^{X}$ and $z_{k}^{X}=\left\vert
\left( -\frac{k_{1}^{X}}{k_{3}^{X}}\right) ^{\frac{1}{k}}\right\vert $,\ $%
z_{kk^{\prime }}^{X}=z_{k}^{X}e^{\frac{2\pi ik^{\prime }}{k}}$,\ $\tilde{z}%
_{kk^{\prime }}^{X}\equiv 1-z_{kk^{\prime }}^{X}$ for $k^{\prime }=0,\cdots
,k-1$. In Eq.(\ref{st}) the D-type Lauricella function $F_{D}^{(K)}$ is one
of the four extensions of the Gauss hypergeometric function to $K$ variables
and is defined as%
\begin{equation}
F_{D}^{(K)}\left( a;b_{1},...,b_{K};c;x_{1},...,x_{K}\right)
=\sum_{n_{1},\cdots ,n_{K}}\frac{\left( a\right) _{n_{1}+\cdots +n_{K}}}{%
\left( c\right) _{n_{1}+\cdots +n_{K}}}\frac{\left( b_{1}\right)
_{n_{1}}\cdots \left( b_{K}\right) _{n_{K}}}{n_{1}!\cdots n_{K}!}%
x_{1}^{n_{1}}\cdots x_{K}^{n_{K}}
\end{equation}%
where $(a)_{n}=a\cdot \left( a+1\right) \cdots \left( a+n-1\right) $ is the
Pochhammer symbol. There was a integral representation of the Lauricella
function $F_{D}^{(K)}$ discovered by Appell and Kampe de Feriet (1926) \cite%
{Appell}%
\begin{align}
& F_{D}^{(K)}\left( a;b_{1},...,b_{K};c;x_{1},...,x_{K}\right)   \notag \\
& =\frac{\Gamma (c)}{\Gamma (a)\Gamma (c-a)}\int_{0}^{1}dt%
\,t^{a-1}(1-t)^{c-a-1}\cdot
(1-x_{1}t)^{-b_{1}}(1-x_{2}t)^{-b_{2}}...(1-x_{K}t)^{-b_{K}}.  \label{Kam}
\end{align}%
The integer $K$ in Eq.(\ref{st}) is defined to be%
\begin{equation}
\text{ }K=\underset{\{\text{for all }p_{j}\neq 0\}}{\sum_{j=1}^{n}j}+%
\underset{\{\text{for all }q_{j}\neq 0\}}{\sum_{j=1}^{m}j}+\underset{\{\text{%
for all }r_{j}\neq 0\}}{\sum_{j=1}^{l}j}.
\end{equation}%
For a given $K$, there can be LSSA with different mass level $N$.

\section{Solving all Lauricella string scattering amplitudes}

To solve all the LSSA, one key simplification of the calculation was the
observation that all arguments $b_{m}$ of the Lauricella functions in the
LSSA are nonpositive integers. We stress that only Lauricella functions with
special arguments are used in the LSSA in Eq.(\ref{st}). As we will see that
this will be the main reason of the solvability of the LSSA.

There seem to be no recurrence relations for the Lauricella functions
available in the literature. We will first generalize the $2+2$ recurrence
relations of the Appell functions to the $K+2$ recurrence relations of the
Lauricella functions. One can then use these $K+2$ recurrence relations to
reduce all the Lauricella functions $F_{D}^{(K)}$ in the LSSA to the Gauss
hypergeometry functions $_{2}F_{1}(a,b,c)$. The next step is to derive a
multiplication theorem for the Gauss hypergeometry functions.

The two results can then be used to prove the solvability of all LSSA. In
the two steps of the proof of the solvability of all the LSSA, the property
of nonpositive integers in the arguments $b_{m}$ of the Lauricella functions
in the LSSA plays a key role in the argument.

\subsection{Recurrence Relations of the LSSA}

We begin with the Appell function case with $K=2$. In the Appell case, there
are four fundamental recurrence relations which link the contiguous
functions
\begin{align}
\left( a-b_{1}-b_{2}\right) F_{1}\left( a;b_{1},b_{2};c,x,y\right)
-aF_{1}\left( a+1;b_{1},b_{2};c,x,y\right) &  \notag \\
+b_{1}F_{1}\left( a;b_{1}+1,b_{2};c,x,y\right) +b_{2}F_{1}\left(
a;b_{1},b_{2}+1;c,x,y\right) & =0, \\
cF_{1}\left( a;b_{1},b_{2};c,x,y\right) -\left( c-a\right) F_{1}\left(
a;b_{1},b_{2};c+1,x,y\right) &  \notag \\
-aF_{1}\left( a+1;b_{1},b_{2};c+1,x,y\right) & =0, \\
cF_{1}\left( a;b_{1},b_{2};c,x,y\right) +c\left( x-1\right) F_{1}\left(
a;b_{1}+1,b_{2};c,x,y\right) &  \notag \\
-\left( c-a\right) xF_{1}\left( a;b_{1}+1,b_{2};c+1,x,y\right) & =0,
\label{c} \\
cF_{1}\left( a;b_{1},b_{2};c,x,y\right) +c\left( y-1\right) F_{1}\left(
a;b_{1},b_{2}+1;c,x,y\right) &  \notag \\
-\left( c-a\right) yF_{1}\left( a;b_{1},b_{2}+1;c+1,x,y\right) & =0.
\label{d}
\end{align}

It is straightforward to generalize the above relations and prove the
following $K+2$ recurrence relations for the $D-type$ Lauricella functions%
\begin{align}
\left( a-\underset{i}{\sum}b_{i}\right) F_{D}^{(K)}\left(
a;b_{1},...,b_{K};c;x_{1},...,x_{K}\right) &  \notag \\
-aF_{D}^{(K)}\left( a+1;b_{1},...,b_{K};c;x_{1},...,x_{K}\right) &  \notag \\
+b_{1}F_{D}^{(K)}\left( a;b_{1}+1,...,b_{K};c;x_{1},...,x_{K}\right) &
\notag \\
+... &  \notag \\
+b_{K}F_{D}^{(K)}\left( a;b_{1},...,b_{K}+1;c;x_{1},...,x_{K}\right) & =0,
\end{align}%
\begin{align}
cF_{D}^{(K)}\left( a;b_{1},...,b_{K};c;x_{1},...,x_{K}\right) &  \notag \\
-\left( c-a\right) F_{D}^{(K)}\left(
a;b_{1},...,b_{K};c+1;x_{1},...,x_{K}\right) &  \notag \\
-aF_{D}^{(K)}\left( a+1;b_{1},...,b_{K};c+1;x_{1},...,x_{K}\right) & =0,
\end{align}%
\begin{align}
cF_{D}^{(K)}\left(
a;b_{1},...,b_{m},...,b_{K};c;x_{1},...,x_{m},...,x_{K}\right) &  \notag \\
+c(x_{m}-1)F_{D}^{(K)}\left(
a;b_{1},...,b_{m}+1,...,b_{K};c;x_{1},...,x_{m},...,x_{K}\right) &  \notag \\
+(a-c)x_{m}F_{D}^{(K)}\left(
a;b_{1},...,b_{m}+1,...,b_{K};c+1;x_{1},...,x_{m},...,x_{K}\right) & =0
\label{NN}
\end{align}
where $m=1,2,...,K.$ One notes that for $K=2$, Eq.(\ref{NN}) reduces to the
Appell recurrence relations in Eq.(\ref{c}) and Eq.(\ref{d}).

To proceed for a fixed $K$, we first introduce two recurrence relations from
Eq.(\ref{NN}) for $m=$ $i$ , $j$ ($i\neq j$)%
\begin{equation}
cF_{D}^{(K)}+c(x_{i}-1)F_{D}^{(K)}\left( b_{i}+1\right)
+(a-c)x_{i}F_{D}^{(K)}\left( b_{i}+1;c+1\right) =0,
\end{equation}%
\begin{equation}
cF_{D}^{(K)}+c(x_{j}-1)F_{D}^{(K)}\left( b_{j}+1\right)
+(a-c)x_{j}F_{D}^{(K)}\left( b_{j}+1;c+1\right) =0
\end{equation}
where in each of the above two equations we have omitted those arguments of $%
F_{D}^{(K)}$ which remain the same for all three Lauricella functions. Then
we shift $b_{i}$ to $b_{i}-1$ and $b_{j}$ to $b_{j}-1$ to obtain%
\begin{equation}
cF_{D}^{(K)}\left( b_{i}-1\right)
+c(x_{i}-1)F_{D}^{(K)}+(a-c)x_{i}F_{D}^{(K)}\left( c+1\right) =0,  \label{i}
\end{equation}%
\begin{equation}
cF_{D}^{(K)}\left( b_{j}-1\right)
+c(x_{j}-1)F_{D}^{(K)}+(a-c)x_{j}F_{D}^{(K)}\left( c+1\right) =0.  \label{j}
\end{equation}
By multiplying $x_{j}$ and $x_{i\text{ }}$in Eq.(\ref{i}) and Eq.(\ref{j})
respectively, we can subtract the resulting two equations to take away the $%
F_{D}^{(K)}\left( c+1\right) $ term and obtain the following key recurrence
relation
\begin{equation}
x_{j}F_{D}^{(K)}\left( b_{i}-1\right) -x_{i}F_{D}^{(K)}\left( b_{j}-1\right)
+\left( x_{i}-x_{j}\right) F_{D}^{(K)}=0.  \label{key}
\end{equation}
One can repeatly apply Eq.(\ref{key}) to the Lauricella functions in the
LSSA in Eq.(\ref{st}) and end up with an expression which expresses $%
F_{D}^{(K)}(b_{1},b_{2},..b_{K})$ in terms of $%
F_{D}^{(K-1)}(b_{1},..b_{i-1},b_{i+1}..b_{j}^{\prime},..b_{K})$, $%
b_{j}^{\prime}=b_{j},b_{j}-1,..,b_{j}-\left\vert b_{i}\right\vert $ or $%
F_{D}^{(K-1)}(b_{1},..b_{i}^{\prime },..b_{j-1},b_{j+1},..b_{K})$, $%
b_{i}^{\prime}=b_{i},b_{i}-1,..,b_{i}-\left\vert b_{j}\right\vert $ (assume $i<j$). We can
repeat the process to decrease the value of $K$ and reduce all the
Lauricella functions $F_{D}^{(K)}$ in the LSSA to the Gauss hypergeometry
functions $F_{D}^{(1)}=$ $_{2}F_{1}(a,b,c,x)$. See Figure 1 in the text.

\begin{figure}[t]
\begin{center}
\includegraphics[
height=2.2in, width=2.8in]
{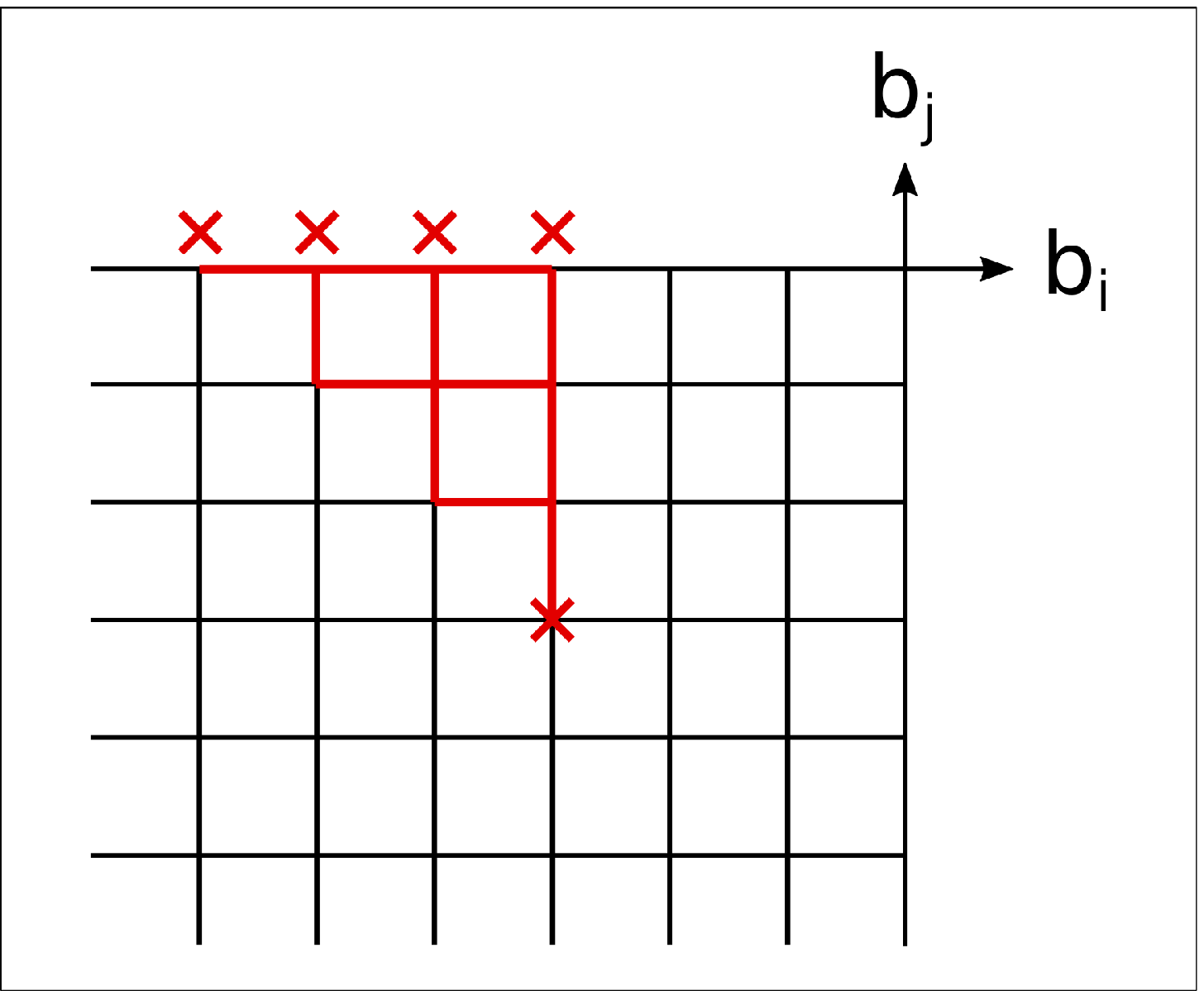} \includegraphics[
height=2.2in, width=2.5in]
{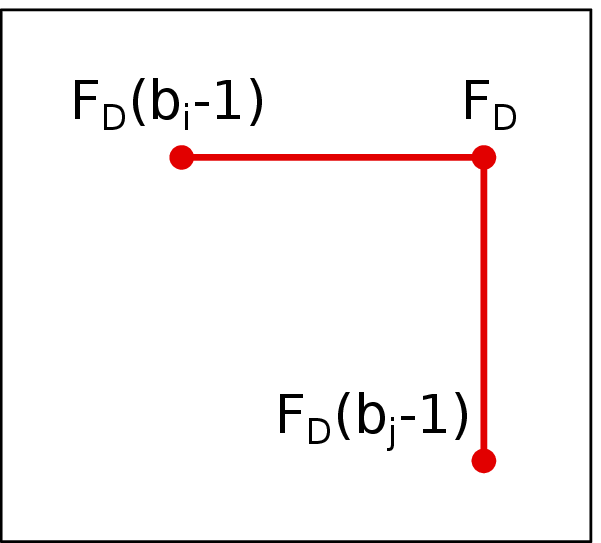} \vskip -0.05cm \hskip -1 cm (a)
\hskip 6.8 cm (b)
\end{center}
\caption{The neighborhood points in the figures are related by the recurrence relations.}%
\label{pic}%
\end{figure}

\subsection{Reduction by a Multiplication Theorem}

In this subsection, to further reduce the Gauss hypergeometry functions in
the LSSA and solve all the LSSA in terms of one single amplitude, we first
derive a multiplication theorem for the Gauss hypergeometry functions.

If we replace $y$ by $(y-1)x$ in the following Taylor's theorem%
\begin{equation}
f(x+y)=\sum_{n=0}^{\infty}f^{(n)}(x)\frac{y^{n}}{n!},
\end{equation}
we get the identity%
\begin{equation}
f(xy)=\sum_{n=0}^{\infty}\frac{(y-1)^{n}x^{n}}{n!}\frac{d^{n}}{dx^{n}}f(x).
\end{equation}
One can now use the derivative relation of the Gauss hypergeometry function%
\begin{equation}
\frac{d^{n}}{dx^{n}}\text{ }_{2}F_{1}(a,b,c,x)=\frac{(a)_{n}(b)_{n}}{(c)_{n}}%
\text{ }_{2}F_{1}(a+n,b+n,c+n,x)
\end{equation}
where $(a)_{n}=a\cdot\left( a+1\right) \cdots\left( a+n-1\right) $ is the
Pochhammer symbol to derive the following multiplication theorem%
\begin{equation}
\text{ }_{2}F_{1}(a,b,c,xy)=\sum_{n=0}^{\left\vert b\right\vert }\frac{%
(y-1)^{n}x^{n}}{n!}\frac{(a)_{n}(b)_{n}}{(c)_{n}}\text{ }%
_{2}F_{1}(a+n,b+n,c+n,x).  \label{M}
\end{equation}
It is important to note that the $\infty$ upper bound of the summation in
the above equation has been replaced by $\left\vert b\right\vert $ since $b$
is a nonpositive integer for the cases of LSSA. In particular if we take $%
x=1 $ in Eq.(\ref{M}), we get the following relation%
\begin{equation}
\text{ }_{2}F_{1}(a,b,c,y)=\sum_{n=0}^{\left\vert b\right\vert }\frac{%
(y-1)^{n}}{n!}\frac{(a)_{n}(b)_{n}}{(c)_{n}}\text{ }_{2}F_{1}(a+n,b+n,c+n,1).
\label{F1}
\end{equation}
The factor $_{2}F_{1}$ on the right hand side of the above equation can be
written as%
\begin{align}
_{2}F_{1}(a+n,b+n,c+n,1) & =\frac{\Gamma(c+n)\Gamma(c-a-b-n)}{\Gamma
(c-a)\Gamma(c-b)}  \notag \\
& =\frac{(-)^{n}(c)_{n}}{(c-a-b)_{n}}\frac{\Gamma(c)\Gamma(c-a-b)}{%
\Gamma(c-a)\Gamma(c-b)}  \notag \\
& =\frac{(-)^{n}(c)_{n}}{(c-a-b)_{n}}\text{ }_{2}F_{1}(a,b,c,1).  \label{F}
\end{align}
Finally one identifies that%
\begin{equation}
_{2}F_{1}(a,b,c,1)=\text{ }_{2}F_{1}(-\frac{t}{2}-1,-p_{1},\frac{u}{2}%
+2-p_{1},1)  \label{F3}
\end{equation}
which corresponds to the LSSA with the polarization $%
(\alpha_{-1}^{T})^{p_{1}}=(\alpha_{-1}^{T})^{-b}$ in Eq.(\ref{p}). One can
even use one of the $15$ Gauss contiguous relations%
\begin{equation}
\{c-2b+(b-a)x\}_{2}F_{1}+b(1-x)_{2}F_{1}(b+1)+(b-c)_{2}F_{1}(b-1)=0,
\label{F4}
\end{equation}
and set $x=1$ which kills the second term of Eq.(\ref{F4}) to reduce $b$ in $%
_{2}F_{1}(a,b,c,1)$ to $b=-1$ or $0$ which corresponds to vector or tachyon
amplitudes in the LSSA. This completes the proof that all the LSSA
calculated in Eq.(\ref{st}) can be solved through various recurrence
relations of Lauricella functions. Moreover, all the LSSA can be expressed
in terms of one single four tachyon amplitude.

For illustration, in the following, we calculate the Lauricella functions
which correspond to the LSSA for levels $K=1,2,3$. For $K=1$ there are three
type of LSSA $(a=-\frac{t}{2}-1,c=\frac{u}{2}+2)$%
\begin{align}
(\alpha_{-1}^{T})^{p_{1}}\text{ , }F_{D}^{(1)}(a,-p_{1},,c-p_{1},1)\text{ , }%
N & =p_{1}, \\
(\alpha_{-1}^{P})^{q_{1}}\text{ , }F_{D}^{(1)}(a,-q_{1},c-q_{1},\left[
\tilde{z}_{1}^{P}\right] )\text{ , }N & =q_{1}, \\
(\alpha_{-1}^{L})^{r_{1}}\text{ , }F_{D}^{(1)}(a,-r_{1},c-r_{1},\left[
\tilde{z}_{1}^{L}\right] )\text{ , }N & =r_{1}.
\end{align}
For $K=2$ there are six type of LSSA $(\omega=-1)$%
\begin{align}
(\alpha_{-1}^{T})^{p_{1}}(\alpha_{-1}^{P})^{q_{1}}\text{,}%
F_{D}^{(2)}(a,-p_{1},-q_{1},c-p_{1}-q_{1},1,\left[ \tilde{z}_{1}^{P}\right] )%
\text{,}N & =p_{1}+q_{1}, \\
(\alpha_{-1}^{T})^{p_{1}}(\alpha_{-1}^{L})^{r_{1}}\text{,}%
F_{D}^{(2)}(a,-p_{1},-r_{1},c-p_{1}-r_{1},1,\left[ \tilde{z}_{1}^{L}\right] )%
\text{,}N & =p_{1}+r_{1}, \\
(\alpha_{-1}^{P})^{q_{1}}(\alpha_{-1}^{L})^{r_{1}}\text{,}%
F_{D}^{(2)}(a,-q_{1},-r_{1},c-q_{1}-r_{1},\left[ \tilde{z}_{1}^{P}\right] ,%
\left[ \tilde{z}_{1}^{L}\right] )\text{,}N & =q_{1}+r_{1}, \\
(\alpha_{-2}^{T})^{p_{2}}\text{ , }F_{D}^{(2)}(a,-p_{2},-p_{2},c-2p_{2},1,1)%
\text{ , }N & =2p_{2}, \\
(\alpha_{-2}^{P})^{q_{2}}\text{ , }%
F_{D}^{(2)}(a,-q_{2},-q_{2},c-2q_{2},1-Z_{2}^{P},1-\omega Z_{2}^{P}),N &
=2q_{2}, \\
(\alpha_{-2}^{L})^{r_{2}}\text{ , }%
F_{D}^{(2)}(a,-r_{2},-r_{2},c-2r_{2},1-Z_{2}^{L},1-\omega Z_{2}^{L}),N &
=2r_{2}.
\end{align}
For $K=3$, there are ten type of LSSA $(\omega_{1}=-1,\omega_{2}=\frac{-1+i%
\sqrt{3}}{2})$%
\begin{align}
(\alpha_{-1}^{T})^{p_{1}}(\alpha_{-1}^{P})^{q_{1}}(%
\alpha_{-1}^{L})^{r_{1}},F_{D}^{(3)}(a,-p_{1},-q_{1},-r_{1},c-p_{1}-q_{1}-r_{1},1,
\left[ \tilde {z}_{1}^{P}\right] ,\left[ \tilde{z}_{1}^{L}\right] ),N &
=p_{1}+q_{1}+r_{1}, \\
(\alpha_{-2}^{T})^{p_{2}}(\alpha_{-1}^{P})^{q_{1}}\text{,}%
F_{D}^{(3)}(a,-p_{2},-p_{2},-q_{1},c-2p_{2}-q_{1},1,1,\left[ \tilde{z}%
_{1}^{P}\right] ),N & =2p_{2}+q_{1}, \\
(\alpha_{-2}^{T})^{p_{2}}(\alpha_{-1}^{L})^{r_{1}}\text{,}%
F_{D}^{(3)}(a,-p_{2},-p_{2},-r_{1},c-2p_{2}-r_{1},1,1,\left[ \tilde{z}%
_{1}^{L}\right] ),N & =2p_{2}+r_{1}, \\
(\alpha_{-1}^{T})^{p_{1}}(\alpha_{-2}^{P})^{q_{2}}\text{,}%
F_{D}^{(3)}(a,-p_{1},-q_{2},-q_{2},c-2q_{2}-p_{1},1,1-Z_{2}^{P},1-%
\omega_{1}Z_{2}^{P}),N & =2q_{2}+p_{1}, \\
(\alpha_{-2}^{P})^{q_{2}}(%
\alpha_{-1}^{L})^{r_{1}},F_{D}^{(3)}(a,-q_{2},-q_{2},-r_{1},c-2q_{2}-r_{1},1-Z_{2}^{P},1-\omega_{1}Z_{2}^{P},
\left[ \tilde{z}_{1}^{L}\right] ),N & =2q_{2}+r_{1}, \\
(\alpha_{-1}^{T})^{p_{1}}(\alpha_{-2}^{L})^{r_{2}}\text{ , }%
F_{D}^{(3)}(a,,-p_{1},-r_{2},-r_{2},c-2r_{2}-p_{1},1,1-Z_{2}^{L},1-%
\omega_{1}Z_{2}^{L}),N & =2r_{2}+p_{1}. \\
(\alpha_{-1}^{P})^{q_{1}}(\alpha_{-2}^{L})^{r_{2}}\text{ , }%
F_{D}^{(3)}(a,,-q_{1},-r_{2},-r_{2},c-2r_{2}-q_{1},\left[ \tilde{z}_{1}^{P}%
\right] ,1-Z_{2}^{L},1-\omega_{1}Z_{2}^{L}),N & =2r_{2}+q_{1}. \\
(\alpha_{-3}^{T})^{p_{3}}\text{ , }%
F_{D}^{(3)}(a,-p_{3},-p_{3},-p_{3},c-3p_{3},1,1,1),N & =3p_{3}, \\
(\alpha_{-3}^{P})^{q_{3}}\text{ , }%
F_{D}^{(3)}(a,-q_{3},-q_{3},-q_{3},c-3q_{3},1-Z_{3}^{P},1-%
\omega_{2}Z_{3}^{P},1-\omega_{2}^{2}Z_{3}^{P}),N & =3q_{3}, \\
(\alpha_{-3}^{L})^{r_{3}}\text{ , }%
F_{D}^{(3)}(a,-r_{3},-r_{3},-r_{3},c-3r_{3},1-Z_{3}^{L},1-%
\omega_{2}Z_{3}^{L},1-\omega_{2}^{2}Z_{3}^{L}),N & =3r_{3}.
\end{align}
All the LSSA for $K=2,3$ can be reduced through the recurrence relations in
Eq.(\ref{key}) and expressed in terms of those of $K=1.$ Furthermore, all
resulting LSSA for $K=1$ can be further reduced by applying Eq.(\ref{F1}) to
Eq.(\ref{F4}) and finally expressed in terms of one single LSSA.

\section{Lauricella Zero Norm States and Ward Identities}

In addition to the recurrence relations among LSSA, there are on-shell
stringy Ward identities among LSSA. These Ward identities can be derived
from the decoupling of two type of zero norm states (ZNS) in the old
covariant first quantized string spectrum. However, as we will see soon that
these Lauricella zero norm states (LZNS) or the corresponding Lauricella
Ward identities are \textit{not} good enough to solve all the LSSA and
express them in terms of one amplitude.

On the other hand, in the last section, we have shown that by using (A)
Recurrence relations of the LSSA, (B) Multiplication theorem of Gauss
hypergeometry function and (C) the explicit calculation of four tachyon
amplitude, one can explicitly solve and calculate all LSSA. This means that
the solvability of LSSA through the calculations of (A), (B) and (C) imply
the validity of Ward identities. Ward identities can not be identities
independent of recurrence relations we used in the last section. Otherwise
there will be a contradiction with the solvabilibity of LSSA.

In this section, we will study some examples of Ward identities of LSSA from
this point of view. Incidentally, high energy zero norm states (HZNS) \cite%
{ChanLee1,ChanLee2,CHL,PRL, CHLTY,susy} and the corresponding stringy Ward
identities at the fixed angle regime, and Regge zero norm states (RZNS) \cite%
{LY,LY2014} and the corresponding Regge Ward identities at the Regge regime
have been studied previously. In particular, HZNS at the fixed angle regime
can be used to solve all the high energy SSA \cite%
{ChanLee1,ChanLee2,CHL,PRL, CHLTY,susy}.

\subsection{The Lauricella zero norm states}

We will consider a smaller set of Ward identities,\ namely, those among the
LSSA or string scattering amplitudes with three tachyons and one arbitrary
string states. So we need only consider polarizations of the tensor states
on the scattering plane. There are two types of zero norm states (ZNS) in
the old covariant first quantum string spectrum

\begin{equation}
\text{Type I}:L_{-1}\left\vert x\right\rangle ,\text{ where }L_{1}\left\vert
x\right\rangle =L_{2}\left\vert x\right\rangle =0,\text{ }L_{0}\left\vert
x\right\rangle =0;  \label{ZN1}
\end{equation}

\begin{equation}
\text{Type II}:(L_{-2}+\frac{3}{2}L_{-1}^{2})\left\vert \widetilde{x}%
\right\rangle ,\text{ where }L_{1}\left\vert \widetilde{x}\right\rangle
=L_{2}\left\vert \widetilde{x}\right\rangle =0,\text{ }(L_{0}+1)\left\vert
\widetilde{x}\right\rangle =0.
\end{equation}
While type I states have zero-norm at any spacetime dimension, type II
states have zero-norm \textit{only} at $D=26$.

We begin with the case of mass level $M^{2}=2$. There is a type II ZNS%
\begin{equation}
\left[ \frac{1}{2}\alpha _{-1}\cdot \alpha _{-1}+\frac{5}{2}k\cdot \alpha
_{-2}+\frac{3}{2}(k\cdot \alpha _{-1})^{2}\right] \left\vert
0,k\right\rangle   \label{2.1}
\end{equation}%
and a type I ZNS

\begin{equation}
\lbrack \theta \cdot \alpha _{-2}+(k\cdot \alpha _{-1})(\theta \cdot \alpha
_{-1})]\left\vert 0,k\right\rangle ,\theta \cdot k=0.  \label{2.2}
\end{equation}%
Note that for the LSSA of three tachyons and one arbitrary string state,
amplitudes with polarizations orthogonal to the scattering plane vanish. We
define the polarizations of the 2nd tensor state with momentum $k_{2}$ on
the scattering plane to be $e_{P}=\frac{1}{M_{2}}(E_{2},\mathrm{k}_{2},0)=%
\frac{k_{2}}{M_{2}}$ $($or $e^{P}=\frac{1}{M_{2}}(-E_{2},\mathrm{k}_{2},0))$
as the momentum polarization, $e^{L}=\frac{1}{M_{2}}(\mathrm{k}_{2},E_{2},0)$
the longitudinal polarization and $e^{T}=(0,0,1)$ the transverse
polarization. $\eta _{\mu \nu }=diag(-1,1,1).$ The three vectors $e^{P}$, $%
e^{L}$ and $e^{T}$ satisfy the completeness relation%
\begin{equation}
\eta _{\mu \nu }=\sum_{\alpha ,\beta }e_{\mu }^{\alpha }e_{\nu }^{\beta
}\eta _{\alpha \beta }
\end{equation}%
where $\mu ,\nu =0,1,2$ and $\alpha ,\beta =P,L,T$ and $\alpha
_{-1}^{T}=\sum_{\mu }e_{\mu }^{T}\alpha _{-1}^{\mu }$, $\alpha
_{-1}^{T}\alpha _{-2}^{L}=\sum_{\mu ,\nu }e_{\mu }^{T}e_{\nu }^{L}\alpha
_{-1}^{\mu }\alpha _{-2}^{\nu }$ etc.

The type II ZNS in Eq.(\ref{2.1}) gives the LZNS%
\begin{equation}
(\sqrt{2}\alpha_{-2}^{P}+\alpha_{-1}^{P}\alpha_{-1}^{P}+\frac{1}{5}\alpha
_{-1}^{L}\alpha_{-1}^{L}+\frac{1}{5}\alpha_{-1}^{T}\alpha_{-1}^{T})|0,k%
\rangle.  \label{R2.3}
\end{equation}
Type I ZNS in Eq.(\ref{2.2}) gives two LZNS%
\begin{equation}
(\alpha_{-2}^{T}+\sqrt{2}\alpha_{-1}^{P}\alpha_{-1}^{T})|0,k\rangle,
\label{R2.1}
\end{equation}%
\begin{equation}
(\alpha_{-2}^{L}+\sqrt{2}\alpha_{-1}^{P}\alpha_{-1}^{L})|0,k\rangle.
\label{R2.2}
\end{equation}
LZNS in Eq.(\ref{R2.1}) and Eq.(\ref{R2.2}) correspond to choose $%
\theta^{\mu }=e^{T}$ and $\theta^{\mu}=e^{L}$ respectively. In conclusion,
there are $3$ LZNS at the mass level $M^{2}=$ $2$.

At the second massive level $M^{2}=4,$ there is a type I scalar ZNS%
\begin{align}
& [\frac{17}{4}(k\cdot\alpha_{-1})^{3}+\frac{9}{2}(k\cdot\alpha_{-1})(%
\alpha_{-1}\cdot\alpha_{-1})+9(\alpha_{-1}\cdot\alpha_{-2})  \notag \\
& +21(k\cdot\alpha_{-1})(k\cdot\alpha_{-2})+25(k\cdot\alpha_{-3})]\left\vert
0,k\right\rangle ,  \label{41}
\end{align}
a symmetric type I spin two ZNS

\begin{equation}
\lbrack 2\theta _{\mu \nu }\alpha _{-1}^{(\mu }\alpha _{-2}^{\nu
)}+k_{\lambda }\theta _{\mu \nu }\alpha _{-1}^{\lambda \mu \nu }]\left\vert
0,k\right\rangle ,k\cdot \theta =\eta ^{\mu \nu }\theta _{\mu \nu }=0,\theta
_{\mu \nu }=\theta _{\nu \mu }  \label{42}
\end{equation}%
where $\alpha _{-1}^{\lambda \mu \nu }\equiv \alpha _{-1}^{\lambda }\alpha
_{-1}^{\mu }\alpha _{-1}^{\nu }$ and two vector ZNS%
\begin{align}
\left[ (\frac{5}{2}k_{\mu }k_{\nu }\theta _{\lambda }^{\prime }+\eta _{\mu
\nu }\theta _{\lambda }^{\prime })\mathcal{\alpha }_{-1}^{(\mu \nu \lambda
)}+9k_{\mu }\theta _{\nu }^{\prime }\mathcal{\alpha }_{-1}^{(\mu \nu
)}+6\theta _{\mu }^{\prime }\mathcal{\alpha }_{-1}^{\mu }\right] \left\vert
0,k\right\rangle ,\theta \cdot k& =0,  \label{43} \\
\left[ (\frac{1}{2}k_{\mu }k_{\nu }\theta _{\lambda }+2\eta _{\mu \nu
}\theta _{\lambda })\mathcal{\alpha }_{-1}^{(\mu \nu \lambda )}+9k_{\mu
}\theta _{\nu }\mathcal{\alpha }_{-1}^{[\mu \nu ]}-6\theta _{\mu }\mathcal{%
\alpha }_{-1}^{\mu }\right] \left\vert 0,k\right\rangle ,\theta \cdot k& =0.
\label{44}
\end{align}%
Note that Eq.(\ref{43}) and Eq.(\ref{44}) are linear combinations of a type
I and a type II ZNS. This completes the four ZNS at the second massive level
$M^{2}=$ $4$.

The scalar ZNS in Eq.(\ref{41}) gives the LZNS%
\begin{equation}
\lbrack25(\alpha_{-1}^{P})^{3}+9\alpha_{-1}^{P}(\alpha_{-1}^{L})^{2}+9%
\alpha_{-1}^{P}(\alpha_{-1}^{T})^{2}+9\alpha_{-2}^{L}\alpha_{-1}^{L}+9%
\alpha_{-2}^{T}\alpha_{-1}^{T}+75\alpha_{-2}^{P}\alpha_{-1}^{P}+50%
\alpha_{-3}^{P}]\left\vert 0,k\right\rangle .  \label{R4.1}
\end{equation}
For the type I spin two ZNS in Eq.(\ref{42}), we define%
\begin{equation}
\theta_{\mu\nu}=\sum_{\alpha,\beta}e_{\mu}^{\alpha}e_{\nu}^{\beta}u_{\alpha%
\beta},
\end{equation}
symmetric and transverse conditions on $\theta_{\mu\nu}$ then implies%
\begin{equation}
u_{\alpha\beta}=u_{\beta\alpha};u_{PP}=u_{PL}=u_{PT}=0.  \label{MM}
\end{equation}
The traceless condition on $\theta_{\mu\nu}$ implies%
\begin{equation}
u_{PP}-u_{LL}-u_{TT}=0.  \label{Naive}
\end{equation}
Eq.(\ref{MM}) and Eq.(\ref{Naive}) give two LZNS%
\begin{equation}
(\alpha_{-1}^{L}\alpha_{-2}^{L}+\alpha_{-1}^{P}\alpha_{-1}^{L}%
\alpha_{-1}^{L}-\alpha_{-1}^{T}\alpha_{-2}^{T}-\alpha_{-1}^{P}%
\alpha_{-1}^{T}\alpha _{-1}^{T})|0,k\rangle,
\end{equation}%
\begin{equation}
(\alpha_{-1}^{(L}\alpha_{-2}^{T)}+\alpha_{-1}^{P}\alpha_{-1}^{L}\alpha
_{-1}^{T})|0,k\rangle.
\end{equation}
The vector ZNS in Eq.(\ref{43}) gives two LZNS%
\begin{equation}
\lbrack6\alpha_{-3}^{T}+18\alpha_{-1}^{(P}\alpha_{-2}^{T)}+9\alpha_{-1}^{P}%
\alpha_{-1}^{P}\alpha_{-1}^{T}+\alpha_{-1}^{L}\alpha_{-1}^{L}\alpha
_{-1}^{T}+\alpha_{-1}^{T}\alpha_{-1}^{T}\alpha_{-1}^{T}]|0,k\rangle,
\end{equation}%
\begin{equation}
\lbrack6\alpha_{-3}^{L}+18\alpha_{-1}^{(P}\alpha_{-2}^{L)}+9\alpha_{-1}^{P}%
\alpha_{-1}^{P}\alpha_{-1}^{L}+\alpha_{-1}^{L}\alpha_{-1}^{L}\alpha
_{-1}^{L}+\alpha_{-1}^{L}\alpha_{-1}^{T}\alpha_{-1}^{T}]|0,k\rangle.
\end{equation}
\ The vector ZNS in Eq.(\ref{44}) gives two LZNS%
\begin{equation}
\lbrack3\alpha_{-3}^{T}-9\alpha_{-1}^{[P}\alpha_{-2}^{T]}-\alpha_{-1}^{L}%
\alpha_{-1}^{L}\alpha_{-1}^{T}-\alpha_{-1}^{T}\alpha_{-1}^{T}\alpha
_{-1}^{T}]|0,k\rangle,
\end{equation}%
\begin{equation}
\lbrack3\alpha_{-3}^{L}-9\alpha_{-1}^{[P}\alpha_{-2}^{L]}-\alpha_{-1}^{L}%
\alpha_{-1}^{L}\alpha_{-1}^{L}-\alpha_{-1}^{L}\alpha_{-1}^{T}\alpha
_{-1}^{T}]|0,k\rangle.
\end{equation}
In conclusion, there are totally $7$ LZNS at the mass level $M^{2}=$ $4$.

It is important to note that there are $9$ LSSA at mass level $M^{2}=$ $2$
with only $3$ LZNS, and $22$ LSSA at mass level $M^{2}=$ $4$ with only $7$
LZNS. So in constrast to the recurrence relations calculated in Eq.(\ref{key}%
) and Eq.(\ref{M}), these Ward identities are not good enough to solve all
the LSSA and express them in terms of one amplitude.

\subsection{The Lauricella Ward identities}

In this subsection, we will explicitly verify some examples of Ward
identities through processes (A),(B) and (C). Process (C) will be implicitly
used through the kinematics. Ward identities can not be identities
independent of recurrence relations we used in processes (A),(B) and (C) in
the last section. We define the following kinematics variables (for $M^{2}=$
$2$)
\begin{equation}
a=\frac{-t}{2}-1=Mk_{3}^{P}-N+1=\sqrt{2}k_{3}^{P}-1,
\end{equation}

\begin{equation}
c=\frac{s}{2}+2-N=-Mk_{1}^{P}=-\sqrt{2}k_{1}^{P},
\end{equation}%
\begin{equation}
d=\left( \frac{-k_{1}^{L}}{k_{3}^{L}}\right) ^{\frac{1}{2}},1-\left( \frac{%
-k_{1}^{P}}{k_{3}^{P}}\right) =\frac{a-c+1}{a+1},
\end{equation}%
then%
\begin{equation}
\frac{u}{2}+2-N=a-c+1-N=a-c-1.
\end{equation}%
\qquad \qquad As the first example, we calculate the Ward identity
associated with the LZNS in Eq.(\ref{R2.1}). The calculation will be based
on processes (A) and (B). By using Eq.(\ref{st}), the Ward identity we want
to prove is%
\begin{align}
& \left( -k_{3}^{T}\right) F_{D}^{(2)}\left( a;-1,-1;a-c-1;1-\left( \frac{%
-k_{1}^{T}}{k_{3}^{T}}\right) ^{\frac{1}{2}},1+\left( \frac{-k_{1}^{T}}{%
k_{3}^{T}}\right) ^{\frac{1}{2}}\right)   \notag \\
& +\sqrt{2}\left( -k_{3}^{P}\right) \left( -k_{3}^{T}\right)
F_{D}^{(2)}\left( a;-1,-1;a-c-1;1-\left( \frac{-k_{1}^{P}}{k_{3}^{P}}\right)
,1-\left( \frac{-k_{1}^{T}}{k_{3}^{T}}\right) \right) =0
\end{align}%
or
\begin{equation}
F_{D}^{(2)}(a;-1,-1;a-c-1;1,1)-(a+1)F_{D}^{(2)}\left( a;-1,-1;a-c-1;\frac{%
a-c+1}{a+1},1\right) =0.  \label{W1}
\end{equation}%
Now let's make use of Eq.(\ref{key}) in the process (A) to the first term of
Eq.(\ref{W1}). We get%
\begin{align}
& 1\times F_{D}^{(2)}(a;-2,0;a-c-1;1,1)  \notag \\
& -1\times F_{D}^{(2)}(a;-1,-1;a-c-1;1,1)  \notag \\
+(1-1)F_{D}^{(2)}(a,;-1,0;a-c-1;1,1)& =0,
\end{align}%
which means
\begin{equation}
F_{D}^{(2)}(a;-1,-1;a-c-1;1)=F_{D}^{(1)}(a;-2;a-c-1;1).
\end{equation}%
Similar calculation can be applied to the second term of Eq.(\ref{W1}),
which can be reduced to%
\begin{align}
& F_{D}^{(2)}\left( a;-1,-1;a-c-1;\frac{a-c+1}{a+1},1\right)   \notag \\
& =\frac{a-c+1}{a+1}F_{D}^{(1)}\left( a;-2;a-c-1;1\right) +\frac{c}{a+1}%
F_{D}^{(1)}\left( a;-1;a-c-1;1\right) .
\end{align}%
Finally the Ward identity in Eq.(\ref{W1}) is explicitly verified through
processe (A)
\begin{align}
& F_{D}^{(2)}(a;-1,-1;a-c-1;1,1)-(a+1)F_{D}^{(2)}\left( a;-1,-1;a-c-1;\frac{%
a-c+1}{a+1},1\right)   \notag \\
& =F_{D}^{(1)}(a;-2;a-c-1;1)-(a+1)  \notag \\
& \times \left[ \frac{a-c+1}{a+1}F_{D}^{(1)}\left( a;-2;a-c-1;1\right) +%
\frac{c}{a+1}F_{D}^{(1)}\left( a;-1;a-c-1;1\right) \right]   \notag \\
& =(c-a)F_{D}^{(1)}\left( a;-2;a-c-1;1\right) -cF_{D}^{(1)}\left(
a;-1;a-c-1;1\right)   \notag \\
& =0
\end{align}%
where Eq.(\ref{F4}) has been used to get the last equality of the above
equation.

As the second example, we calculate the Ward identity associated with the
LZNS in Eq.(\ref{R2.2}). By using Eq.(\ref{st}), the Ward identity is%
\begin{align}
& \left( -k_{3}^{L}\right) F_{D}^{(2)}\left( a;-1,-1;a-c-1;1-\left( \frac{%
-k_{1}^{L}}{k_{3}^{L}}\right) ^{\frac{1}{2}},1+\left( \frac{-k_{1}^{L}}{%
k_{3}^{L}}\right) ^{\frac{1}{2}}\right)   \notag \\
& +\sqrt{2}\left( -k_{3}^{P}\right) \left( -k_{3}^{L}\right)
F_{D}^{(2)}\left( a;-1,-1;a-c-1;1-\left( \frac{-k_{1}^{P}}{k_{3}^{P}}\right)
,1-\left( \frac{-k_{1}^{L}}{k_{3}^{L}}\right) \right) =0
\end{align}%
or
\begin{align}
& F_{D}^{(2)}(a;-1,-1;a-c-1;1-d,1+d)  \notag \\
& -(a+1)F_{D}^{(2)}\left( a;-1,-1;a-c-1;\frac{a-c+1}{a+1},1-d^{2}\right) =0.
\label{W2}
\end{align}%
Now let's make use of Eq.(\ref{key}) in the process (A) to the first term of
Eq.(\ref{W2}). We get%
\begin{align}
& F_{D}^{(2)}F(a;-1,-1;a-c-1;1-d,1+d)  \notag \\
& =\frac{1-d}{1+d}F_{D}^{(1)}(a;-2;a-c-1;1+d)-\frac{2d}{1+d}%
F_{D}^{(1)}(a;-1;a-c-1;1+d).  \label{WW2}
\end{align}%
We then use Eq.(\ref{M}) and Eq.(\ref{F}) in the prosess (B) to simplify the
first term and the second term on the r.h.s. of Eq.(\ref{WW2}) to be%
\begin{equation}
F_{D}^{(1)}(a;-2;a-c-1;1+d)=F_{D}^{(1)}(a;-2;a-c-1;1)\left[ 1-\frac{2ad}{c-1}%
+\frac{a(a+1)^{2}}{(c-1)(c-2)}\right]   \notag
\end{equation}%
and%
\begin{equation}
F_{D}^{(1)}(a;-1;a-c-1;1+d)=F_{D}^{(1)}(a;-1;a-c-1;1)\left[ 1-\frac{ad}{c}%
\right] .  \notag
\end{equation}%
We now use Eq.(\ref{key}) in the process (A) to the second term of Eq.(\ref%
{W2}) to get%
\begin{align}
& F_{D}^{(2)}\left( a;-1,-1;a-c-1;\frac{a-c+1}{a+1},1-d^{2}\right)   \notag
\\
& =\left[ \frac{a-c+1}{(a+1)(1-d^{2})}\right] F_{D}^{(1)}\left(
a;-2;a-c-1;1-d^{2}\right)   \notag \\
& -\left[ \frac{a-c+1}{(a+1)(1-d^{2})}-(1-d)\right]
F_{D}^{(1)}F(a;-1;a-c-1;1-d^{2})  \label{WW3}
\end{align}%
We then use Eq.(\ref{M}) and Eq.(\ref{F}) in the prosess (B) to simplify the
first term and the second term on the r.h.s. of Eq.(\ref{WW3}) to be%
\begin{align}
& F_{D}^{(1)}(a;-2;a-c-1;1-d^{2})  \notag \\
& =\left[ 1+\frac{2ad^{2}}{c-1}+\frac{a(a+1)d^{4}}{(c-1)(c-2)}\right]
F_{D}^{(1)}(a;-2;a-c-1,1)
\end{align}%
and%
\begin{align}
& F_{D}^{(1)}(a;-1;a-c-1;1-d^{2})  \notag \\
& =\left( 1+\frac{ad^{2}}{c}\right) F_{D}^{(1)}F(a;-1;a-c-1,1).
\end{align}%
Finally we put all results to Eq.(\ref{W2}) and end up with%
\begin{align}
& F_{D}^{(2)}(a;-1,-1;a-c-1;1-d,1+d)  \notag \\
& -(a+1)F_{D}^{(2)}\left( a;-1,-1;a-c-1;\frac{a-c+1}{a+1},1-d^{2}\right)
\notag \\
& =\frac{1-d}{1+d}F_{D}^{(1)}(a;-2;a-c-1;1+d)  \notag \\
& -\frac{2d}{1+d}F_{D}^{(1)}(a;-1;a-c-1,1+d)  \notag \\
& -(a+1)\left[ \frac{a-c+1}{(a+1)(1-d^{2})}\right] F_{D}^{(1)}\left(
a;-2;a-c-1;1-d^{2}\right)   \notag \\
& +(a+1)\left[ \frac{a-c+1}{(a+1)(1-d^{2})}-(1-d)\right]
F_{D}^{(1)}(a;-1;a-c-1;1-d^{2})  \notag \\
& =\frac{1-d}{1+d}\left[ 1-\frac{2ad}{c-1}+\frac{a(a+1)^{2}}{(c-1)(c-2)}%
\right] F_{D}^{(1)}(a;-2;a-c-1;1)  \notag \\
& -\frac{2d}{1+d}\left[ 1-\frac{ad}{c}\right] F_{D}^{(1)}(a;-1;a-c-1;1)
\notag \\
& -(a+1)\left[ \frac{a-c+1}{(a+1)(1-d^{2})}\right] \left[ 1+\frac{2ad^{2}}{%
c-1}+\frac{a(a+1)d^{4}}{(c-1)(c-2)}\right] F_{D}^{(1)}(a;-2;a-c-1;1)  \notag
\\
& +(a+1)\left[ \frac{a-c+1}{(a+1)(1-d^{2})}-(1-d)\right] \left( 1+\frac{%
ad^{2}}{c}\right) F_{D}^{(1)}(a;-1;a-c-1;1)  \notag \\
& =0
\end{align}%
where again Eq.(\ref{F4}) has been used to get the last equality of the
above equation.

\section{Conclusions}

In this paper we have shown that there exist infinite number of recurrence
relations valid for \textit{all} energies among the LSSA of three tachyons
and one arbitrary string state. Moreover, these infinite number of
recurrence relations can be used to solve all the LSSA and express them in
terms of one single four tachyon amplitude. It will be interesting to see
the relationship between the solvability of LSSA and the associated $%
SL(K+3,C)$ \cite{sl5c,sl4c} symmetry of the Lauricella functions as
suggested in \cite{LLY2}. The results of this calculation extend the
solvability of SSA at the high energy, fixed angle scattering limit \cite%
{ChanLee1,ChanLee2,CHL,PRL, CHLTY,susy} and those at the Regge scattering
limit \cite{LY,LY2014} discovered previously.

We have also shown that for the first few mass levels the solvability of
LSSA through the calculations of recurrence relations implies the validity
of Ward identities derived from the decoupling of LZNS. However the
Lauricella Ward identities are \textit{not} good enough to solve all the
LSSA and express them in terms of one amplitude.

The extention of results for the one tensor three tachyon scatterings
calculated in this paper to multi-tensor scatterings is currently under
investigation.

\vskip 1cm

\begin{acknowledgments}
The works of JCL and YY are supported in part by the Ministry of Science and
Technology and S.T. Yau center of NCTU, Taiwan. TL was supported by research
grant 2017 of Kangwon National University.
\end{acknowledgments}


\end{document}